\documentclass[prl,showpacs,preprintnumbers]{revtex4}
\usepackage{amsfonts}
\usepackage{mdframed}
\usepackage{amssymb}
\usepackage{footnote}
\usepackage{amsmath}
\usepackage{graphicx}
\usepackage{float}
\usepackage[font={footnotesize,it}]{caption}
\usepackage[utf8]{inputenc}
\usepackage{natbib}
\usepackage{tikz}
\usepackage{tikz-3dplot}
\usepackage[colorlinks=true,
            linkcolor=purple,
            urlcolor=purple,
            citecolor=blue]{hyperref}

\begin{document}

\title{Quantum particle on a surface: Catenary surface and paraboloid of
revolution}
\author{S. Habib Mazharimousavi}
\email{habib.mazhari@emu.edu.tr}
\affiliation{Department of Physics, Faculty of Arts and Sciences, Eastern Mediterranean
University, Famagusta, North Cyprus via Mersin 10, Turkey}
\date{\today }

\begin{abstract}
We revisit the Schr\"{o}dinger equation of a quantum particle that is
confined on a curved surface. Inspired by the novel work of R. C. T. da
Costa [1] we find the field equation in a more convenient notation. The
contribution of the principal curvatures in the effective binding potential
on the surface is emphasized. Furthermore, using the so-called Monge-Gauge
we construct the approximate Schr\"{o}dinger equation for a flat surface
with small fluctuations. Finally, the resulting Schr\"{o}dinger equation is
solved for some specific surfaces. In particular, we give exact solutions
for a particle confined on a Catenary surface and a paraboloid of revolution.
\end{abstract}

\maketitle

\section{Introduction}

Nowadays, studying the propagation of quantum particles on curved surfaces
became of interest in different areas of experimental and theoretical
physics. Graphene with only one atom thickness is one of such 2-dimensional
surface-like materials \cite{Graphene}. Furthermore, another surface-like
material is the so-called carbon nanotube \cite{CNT} with stronger $sp^{2}$
bonds which make it one of the best thermal conductors \cite{CNT2}. Another
important 2D material that has been the subject of intensive research both
experimentally and theoretically is Phosphorene, a monolayer of black
phosphorus \cite{PHOS} (see also the references therein). It is a 2D
semiconductor material with an anisotropic orthorhombic structure and high
optical and UV absorption. To this list of 2-dimensional materials we would
like to add the so-called fluid lipid membranes \cite{FLM} which has shown a
growing interest among physicists, mathematicians, and biologists.

Concerning these highly important lower-dimensional materials, and the
quantum phenomenon on these surfaces, one has to create/construct an induced
lower-dimensional quantum mechanics which may be applied to such
2-dimensional materials. Such a formalism has been introduced long time ago
in \cite{2dQ} and \cite{2dQ2}. However, the recent work of Ferrari and
Cuoghi \cite{PRL1}, proves that the story has not been over yet. To
summarize the difference between these three works, one observes the
following. In \cite{2dQ}, the action principle has been considered directly
in n-dimensional curved space. Hence, considering $n=2$ one gets the quantum
on a curved surface. In \cite{2dQ2}, using the differential geometrical
properties of the curved surface, the Schr\"{o}dinger equation in
3-dimensional Euclidean space has been reduced to the Schr\"{o}dinger
equation on the surface. In \cite{PRL1}, the formalism of \cite{2dQ2} has
been extended by including the electric and magnetic fields. Following these
seminal works, there have been significant improvements in the quantum
systems in the two-dimensional curved surfaces as well as on curves. In this
line, we may refer to \cite{Flipe} where the effects of the geometry, as
well as magnetic field on the electronic transport properties of metallic
nanotubes, have been numerically investigated. In \cite{Jose}, an electron
confined on a torus under the influence of external electric and magnetic
fields has been numerically studied. In \cite{Silva} the authors studied the
an electron on a catenoid surface. Oliveira et al in \cite{Oliveria} solved
the Schr\"{o}dinger equation on a sphere under non-central potential and
Schmidt in \cite{Schmidt} and \cite{Schmidt2} introduced exact solutions of
Schr\"{o}dinger equation for a charged particle confined on a sphere, on a
cylinder and on a torus while is imposed with uniform electric and magnetic
fields. Furthermore, electrons confined on a rotating sphere in the presence
of a magnetic field have been considered by Lima et al in \cite{Lima}, and
the effects of the rotation were compared with the effects of the magnetic
fields. The application of this formalism has not been limited only to these
works, for instance, we refer to \cite{Application} for further reading.

\subsection{Our motivation}

We observed that most of the papers published recently are based on the
effective Schr\"{o}dinger equation derived in \cite{2dQ2}, particularly Eq.
(14). In driving this equation, R. C. T. da Costa used a kind of unfamiliar
notation to our new generation of young physicists. For instance, while
these days we are very careful on distinguishing between contravariant and
covariant vectors especially when the contraction of tensors is in the
subject, in \cite{2dQ2}, it was only a matter of notation. Therefore, many
steps in finding the effective Schr\"{o}dinger equation in \cite{2dQ2} are
unfamiliar. The aim of this paper is first to construct a full detailed
calculation with modern - so to say - notation toward the effective Schr\"{o}%
dinger equation. We should add that there are some other works that looked
at this issue from another perspective. For instance, a very general, as
well as an interesting approach to the constraint motion of a quantum
particle in n-dimensional Euclidian space, has been studied by P.C. Schuster
and R.L. Jaffe in \cite{Schuster}. In addition, the applications of some
specific parametrization such as Monge parametrization seem to be missing in
the literature. Since for surfaces such as graphene, the small deviation
from a flat surface may be of interest the application of Monge gauge
becomes important. Hence, we study the quantum particle under Monge
parametrization and for small deviation deviations, we present the
simplified approximate effective Schr\"{o}dinger equation. Finally, the
exact solutions of the effective Schr\"{o}dinger equation for quantum
particles confined on some important curved surfaces such as Catenoid and
paraboloid of revolution are missing. Therefore, we investigate the possible
exact solutions for these two cases.

\section{Schr\"{o}dinger equation on a curved surface}

We consider a quantum particle of mass $m$ confined on a differentiable
surface $S$ in the three-dimensional Euclidean space. We also adopt a local
two-dimensional coordinate system $\left\{ u^{1},u^{2}\right\} $ on the
surface and a mapping $\mathbf{r:}=\mathbf{r}\left( u^{1},u^{2}\right) $
which assigns any point on the surface to a point on the three-dimensional
space, i.e.,%
\begin{equation}
\mathbf{r:}=\mathbf{r}\left( u^{1},u^{2}\right) =\left( x^{1}\left(
u^{1},u^{2}\right) ,x^{2}\left( u^{1},u^{2}\right) ,x^{3}\left(
u^{1},u^{2}\right) \right) .
\end{equation}%
Herein, the position vector $\mathbf{r}=\left( x^{1},x^{2},x^{3}\right) $
represents a point on the surface while a point in the neighborhood of the
surface can be described using an additional coordinate i.e., $u^{3}$ in the
direction normal to the surface (see Fig. 1).

\begin{figure}[tbp]
\caption{\protect\includegraphics[scale=0.8]{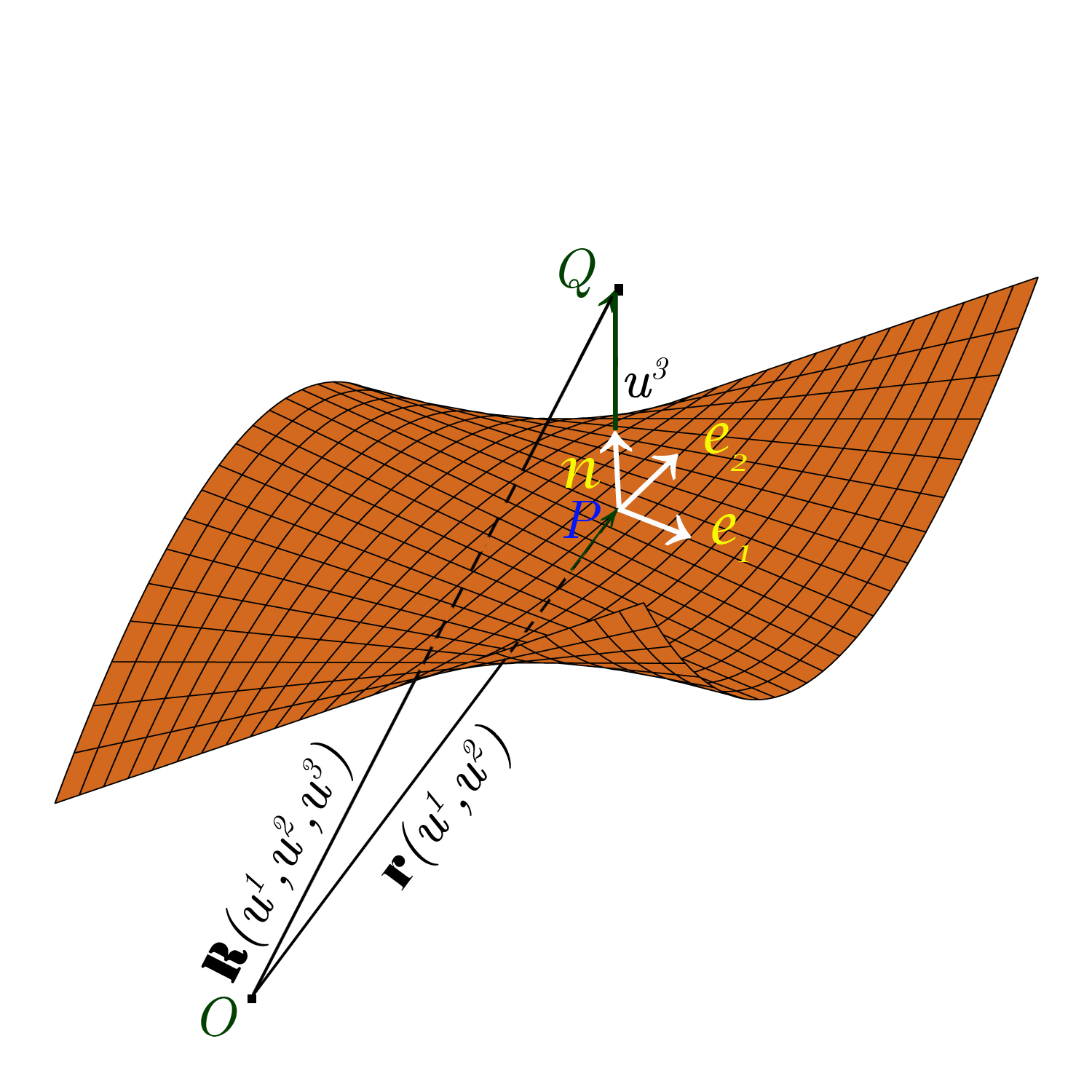}}
The curvilinear coordinate system addressing the surface and the ambience.
\end{figure}

Hence, one writes%
\begin{equation}
\mathbf{R}\left( u^{1},u^{2},u^{3}\right) =\mathbf{r}\left(
u^{1},u^{2}\right) +u^{3}\mathbf{n.}
\end{equation}%
We note that the tangent vectors 
\begin{equation}
\mathbf{e}_{a}=\frac{\partial \mathbf{r}}{\partial u^{a}}=\mathbf{\partial }%
_{a}\mathbf{r}
\end{equation}%
with $a=1,2,$ make a local $2$-dimensional coordinate system that spans the
tangent surface to $S$ at any point $P$ on $S$ where $\mathbf{e}_{a}$ are
determined. Furthermore, the three-dimensional coordinate system consists of
two tangent vectors $\mathbf{e}_{a}$ and the unit normal $\mathbf{n}$ make a
local 3-dimensional coordinate system which describes the space surrounding
the surface $S$ such that%
\begin{equation}
\mathbf{n}=\frac{\mathbf{e}_{1}\times \mathbf{e}_{2}}{\left\vert \mathbf{e}%
_{1}\times \mathbf{e}_{2}\right\vert }.
\end{equation}

We add that, in general, $\mathbf{e}_{a}$ are neither unit vectors nor
orthogonal, however, $\mathbf{n}$ is unit vector and normal to $\mathbf{e}%
_{a}.$ Furthermore, the $3$-dimensional coordinate system as well as the two
dimensional one are curvilinear, and based on the tangent vectors $\mathbf{e}%
_{a}$ and the normal vector $\mathbf{n}$, one constructs the metric tensor
of the space and the surface as defined by%
\begin{equation}
g_{\mu \nu }=\frac{\partial \mathbf{R}}{\partial u^{\mu }}.\frac{\partial 
\mathbf{R}}{\partial u^{\nu }}
\end{equation}%
and%
\begin{equation}
h_{ab}=\frac{\partial \mathbf{r}}{\partial u^{a}}.\frac{\partial \mathbf{r}}{%
\partial u^{b}},
\end{equation}%
respectively, in which $\mu ,\nu =1,2,3$ and $a,b=1,2.$ Let's add that $%
h_{ab}$ is called the first fundamental form of the surface $S$ which is an
intrinsic geometrical property of the surface. In addition to the first
fundamental form, there exist the second fundamental form of the surface
which is an extrinsic property of the surface and is defined by%
\begin{equation}
k_{ab}:=\mathbf{e}_{a}.\partial _{b}\mathbf{n}
\end{equation}%
which due to the fact that $\mathbf{e}_{a}.\mathbf{n}=0$, it is also equal to%
\begin{equation}
k_{ab}=-\mathbf{n}.\partial _{ab}\mathbf{r.}
\end{equation}

The second fundamental form $k_{ab}$ is also called the extrinsic curvature
tensor because it is defined in terms of the normal vector $\mathbf{n}$
which is an indication of the embedding of the surface in a
higher-dimensional space/ambiance..

Coming back to the metric tensor of the space surrounding the surface one
writes%
\begin{equation}
g_{ab}=\frac{\partial \mathbf{R}}{\partial u^{a}}.\frac{\partial \mathbf{R}}{%
\partial u^{b}}=\left( \mathbf{e}_{a}+u^{3}\partial _{a}\mathbf{n}\right)
.\left( \mathbf{e}_{b}+u^{3}\partial _{b}\mathbf{n}\right) ,  \label{MT}
\end{equation}%
\begin{equation}
g_{a3}=g_{3a}=\left( \mathbf{e}_{a}+u^{3}\partial _{a}\mathbf{n}\right) .%
\mathbf{n}
\end{equation}%
and%
\begin{equation}
g_{33}=\mathbf{n}.\mathbf{n}=1.
\end{equation}

To calculate $g_{ab},$ $g_{a3}$ and $g_{3a}$ we apply the so-called equation
of Weingarten which states%
\begin{equation}
\partial _{a}\mathbf{n}=k_{a}^{b}\mathbf{e}_{b}  \label{WE}
\end{equation}%
in which $k_{a}^{b}=h^{bc}k_{ac}$ is the mixed form of the extrinsic
curvature tensor and $h^{bc}$ is the inverse of the metric tensor $h_{bc}$
such that $h_{ac}h^{cb}=\delta _{a}^{b}.$ Considering (\ref{WE}) in (\ref{MT}%
) one finds%
\begin{equation}
g_{ab}=\mathbf{e}_{a}.\mathbf{e}_{b}+u^{3}\partial _{a}\mathbf{n.e}_{b}+u^{3}%
\mathbf{e}_{a}.\partial _{b}\mathbf{n+}\left( u^{3}\right) ^{2}\partial _{a}%
\mathbf{n.}\partial _{b}\mathbf{n}
\end{equation}%
or simply%
\begin{equation}
g_{ab}=h_{ab}+u^{3}k_{a}^{c}h_{cb}+u^{3}k_{b}^{c}h_{ac}\mathbf{+}\left(
u^{3}\right) ^{2}k_{a}^{c}k_{b}^{d}h_{cd}
\end{equation}%
and $g_{a3}=g_{3a}=0.$ Having the first and the second fundamental forms
symmetric, we obtain%
\begin{equation}
g_{ab}=h_{ab}+2u^{3}k_{a}^{c}h_{cb}\mathbf{+}\left( u^{3}\right)
^{2}k_{a}^{c}k_{b}^{d}h_{cd}.
\end{equation}

Using the so-called Laplace-Beltrami operator the Schr\"{o}dinger equation
of a particle in the space spanned with $g_{\mu \nu }$ is given by%
\begin{equation}
-\frac{\hslash ^{2}}{2m}\frac{1}{\sqrt{g}}\partial _{\mu }\left( \sqrt{g}%
g^{\mu \nu }\partial _{\nu }\right) \psi \left( u^{\alpha },t\right) +V_{0}%
\tilde{\delta}\left( u^{3}\right) \psi \left( u^{\alpha },t\right) =i\hslash 
\frac{\partial }{\partial t}\psi \left( u^{\alpha },t\right)   \label{SE}
\end{equation}%
in which $V_{0}\tilde{\delta}\left( u^{3}\right) $ is a potential which
confines the particle on the surface $S$ and $\tilde{\delta}\left(
u^{3}\right) $ is an anti-Dirac delta function such that%
\begin{equation}
\tilde{\delta}\left( u^{3}\right) =\left\{ 
\begin{array}{cc}
0 & u^{3}=0 \\ 
\infty  & u^{3}\neq 0%
\end{array}%
\right. .
\end{equation}

To simplify (\ref{SE}), one needs to calculate $g=\det g_{\mu \nu }$, i.e.%
\begin{equation}
g=\det \left[ 
\begin{array}{ccc}
g_{11} & g_{12} & 0 \\ 
g_{21} & g_{22} & 0 \\ 
0 & 0 & 1%
\end{array}%
\right] =\det \left[ 
\begin{array}{cc}
g_{11} & g_{12} \\ 
g_{21} & g_{22}%
\end{array}%
\right] =g_{11}g_{22}-\left( g_{12}\right) ^{2}
\end{equation}%
where%
\begin{equation}
g_{11}=h_{11}+2u^{3}k_{1}^{c}h_{c1}\mathbf{+}\left( u^{3}\right)
^{2}k_{1}^{c}k_{1}^{d}h_{cd}
\end{equation}%
\begin{equation}
g_{22}=h_{22}+2u^{3}k_{2}^{c}h_{c2}\mathbf{+}\left( u^{3}\right)
^{2}k_{2}^{c}k_{2}^{d}h_{cd}
\end{equation}%
and%
\begin{equation}
g_{12}=g_{21}=h_{12}+2u^{3}k_{1}^{c}h_{c2}\mathbf{+}\left( u^{3}\right)
^{2}k_{1}^{c}k_{2}^{d}h_{cd}
\end{equation}%
upon which 
\begin{multline}
g=\left( h_{11}+2u^{3}k_{1}^{c}h_{c1}\mathbf{+}\left( u^{3}\right)
^{2}k_{1}^{c}k_{1}^{d}h_{cd}\right) \left( h_{22}+2u^{3}k_{2}^{a}h_{a2}%
\mathbf{+}\left( u^{3}\right) ^{2}k_{2}^{a}k_{2}^{b}h_{ab}\right) - \\
\left( h_{12}+2u^{3}k_{1}^{c}h_{c2}\mathbf{+}\left( u^{3}\right)
^{2}k_{1}^{c}k_{2}^{d}h_{cd}\right) \left( h_{21}+2u^{3}k_{2}^{a}h_{a1}%
\mathbf{+}\left( u^{3}\right) ^{2}k_{2}^{a}k_{1}^{b}h_{ab}\right) .
\end{multline}

After some manipulation, the latter equation reduces to%
\begin{equation}
g=\left( h_{22}h_{11}-h_{12}^{2}\right) \left( 1+u^{3}\left(
k_{1}^{1}+k_{2}^{2}\right) +\left( u^{3}\right) ^{2}\left(
k_{1}^{1}k_{2}^{2}-k_{1}^{2}k_{2}^{1}\right) \right) ^{2}.
\end{equation}%
We remember that the trace and the determinant of the extrinsic curvature
tensor are invariant under the coordinate transformation. Therefore,
although $k_{a}^{b}$ is not diagonal and consequently $k_{1}^{1}$ and $%
k_{2}^{2}$ are not the principal curvatures, but $k_{1}^{1}+k_{2}^{2}=Tr%
\left( k_{a}^{b}\right) =$ $\tilde{k}_{1}^{1}+\tilde{k}_{2}^{2}$ and $%
k_{1}^{1}k_{2}^{2}-k_{1}^{2}k_{2}^{1}=\det \left( k_{a}^{b}\right) =\tilde{k}%
_{1}^{1}\check{k}_{2}^{2}$ in which $\tilde{k}_{1}^{1}$ and $\tilde{k}%
_{2}^{2}$ are the principal curvatures of the surface. Considering the above
facts one writes 
\begin{equation}
g=h\left( 1+u^{3}\left( \tilde{k}_{1}^{1}+\tilde{k}_{2}^{2}\right) +\left(
u^{3}\right) ^{2}\tilde{k}_{1}^{1}\check{k}_{2}^{2}\right) ^{2}  \label{D}
\end{equation}%
where 
\begin{equation}
h=h_{22}h_{11}-h_{12}^{2}=\det \left( h_{ab}\right) .
\end{equation}

We also recall the definition of the Gaussian and total curvatures in terms
of the  principal curvatures which are defined as%
\begin{equation}
K_{G}=\tilde{k}_{1}^{1}\check{k}_{2}^{2}
\end{equation}%
and%
\begin{equation}
K=\tilde{k}_{1}^{1}+\tilde{k}_{2}^{2},
\end{equation}%
respectively, which simplifies (\ref{D}) as%
\begin{equation}
g=h\left( 1+u^{3}K+\left( u^{3}\right) ^{2}K_{G}\right) ^{2}.  \label{DMT}
\end{equation}%
Next, we substitute (\ref{DMT}) into the Schr\"{o}dinger equation (\ref{SE})
to get%
\begin{equation*}
-\frac{\hslash ^{2}}{2m}\frac{1}{\omega \sqrt{h}}\partial _{a}\left( \omega 
\sqrt{h}g^{ab}\partial _{b}\right) \psi \left( u^{\alpha },t\right) -\frac{%
\hslash ^{2}}{2m}\frac{1}{\omega }\partial _{3}\left( \omega \partial
_{3}\right) \psi \left( u^{\alpha },t\right) +V_{0}\tilde{\delta}\left(
u^{3}\right) \psi \left( u^{\alpha },t\right) =i\hslash \frac{\partial }{%
\partial t}\psi \left( u^{\alpha },t\right) 
\end{equation*}%
where $\omega =1+u^{3}K+\left( u^{3}\right) ^{2}K_{G}.$

Introducing $\psi \left( u^{\alpha },t\right) =\frac{1}{\sqrt{\omega }}\psi
_{t}\left( u^{a}\right) \psi _{n}\left( u^{3}\right) e^{-iEt/\hslash }$ one
obtains 
\begin{equation*}
-\frac{\hslash ^{2}}{2m}\frac{1}{\sqrt{\omega }\sqrt{h}\psi _{t}\left(
u^{a}\right) }\partial _{a}\left( \omega \sqrt{h}g^{ab}\partial _{b}\right) 
\frac{1}{\sqrt{\omega }}\psi _{t}\left( u^{a}\right) -\frac{\hslash ^{2}}{2m}%
\frac{1}{\sqrt{\omega }\psi _{n}\left( u^{3}\right) }\partial _{3}\left(
\omega \partial _{3}\right) \frac{1}{\sqrt{\omega }}\psi _{n}\left(
u^{3}\right) +V_{0}\tilde{\delta}\left( u^{3}\right) =E
\end{equation*}%
which after some manipulation becomes%
\begin{multline}
-\frac{\hslash ^{2}}{2m}\frac{1}{\sqrt{\omega }\sqrt{h}\psi _{t}\left(
u^{a}\right) }\partial _{a}\left( \omega \sqrt{h}g^{ab}\partial _{b}\right) 
\frac{1}{\sqrt{\omega }}\psi _{t}\left( u^{a}\right) - \\
\frac{\hslash ^{2}}{2m}\frac{1}{\psi _{n}\left( u^{3}\right) }\left( \psi
_{n}^{\prime \prime }\left( u^{3}\right) +\left( \left( \frac{\partial
_{3}\omega }{2\omega }\right) ^{2}-\frac{\partial _{3}^{2}\omega }{2\omega }%
\right) \psi _{n}\left( u^{3}\right) \right) +V_{0}\tilde{\delta}\left(
u^{3}\right) =E.
\end{multline}

Considering the fact that, on the surface where the particle is confined, $%
u^{3}=0$ we obtain $\omega =1,$ $\partial _{3}\omega =K,$ and $\partial
_{3}^{2}\omega =2K_{G\text{ }}.$ Hence, (29) becomes%
\begin{equation}
-\frac{\hslash ^{2}}{2m}\frac{1}{\sqrt{h}\psi _{t}\left( u^{a}\right) }%
\partial _{a}\left( \sqrt{h}g^{ab}\partial _{b}\right) \psi _{t}\left(
u^{a}\right) -\frac{\hslash ^{2}}{2m}\left( \frac{\psi _{n}^{\prime \prime
}\left( u^{3}\right) }{\psi _{n}\left( u^{3}\right) }+\left( \left( \frac{K}{%
2}\right) ^{2}-K_{G}\right) \right) +V_{0}\tilde{\delta}\left( u^{3}\right)
=E.
\end{equation}%
After separating the equation to "on the surface" and "normal to the
surface" one gets%
\begin{equation}
-\frac{\hslash ^{2}}{2m}\frac{1}{\sqrt{h}}\partial _{a}\left( \sqrt{h}%
g^{ab}\partial _{b}\right) \psi _{t}\left( u^{a}\right) -\frac{\hslash ^{2}}{%
2m}\left( \left( \frac{K}{2}\right) ^{2}-K_{G}\right) \psi _{t}\left(
u^{a}\right) =E_{t}\psi _{t}\left( u^{a}\right) 
\end{equation}%
and%
\begin{equation}
-\frac{\hslash ^{2}}{2m}\psi _{n}^{\prime \prime }\left( u^{3}\right) +V_{0}%
\tilde{\delta}\left( u^{3}\right) \psi _{n}\left( u^{3}\right) =E_{n}\psi
_{n}\left( u^{3}\right) ,
\end{equation}%
respectively. The first equation is the two-dimensional Schr\"{o}dinger
equation of the particle on the surface and the second equation is the
one-dimensional Schr\"{o}dinger equation normal to the surface. The total
energy of the particle is given by $E=E_{t}+E_{n}.$

In this study, we consider 
\begin{equation}
\tilde{\delta}\left( u^{3}\right) =\left\{ 
\begin{array}{cc}
0 & 0<u^{3}<\epsilon \\ 
\infty & \text{elsewhere}%
\end{array}%
\right.
\end{equation}%
which yields%
\begin{equation}
\psi _{n}\left( u^{3}\right) =\sqrt{\frac{\epsilon }{2}}\sin \left( \frac{%
\nu \pi }{\epsilon }u^{3}\right)
\end{equation}%
with energy 
\begin{equation}
\left( E_{n}\right) _{\nu }=\frac{\nu ^{2}\pi ^{2}\hslash ^{2}}{2m\epsilon
^{2}}
\end{equation}%
where $\nu =1,2,3,...$ and $\epsilon $ is the thickness of the surface.

On the other hand, the tangent Schr\"{o}dinger equation implies a
non-positive effective potential which is purely geometric., i.e.%
\begin{equation}
V_{S}=-\frac{\hslash ^{2}}{2m}\left( \left( \frac{K}{2}\right)
^{2}-K_{G}\right) 
\end{equation}%
which in terms of the principal curvatures of the surface becomes%
\begin{equation}
V_{S}=-\frac{\hslash ^{2}}{8m}\left( \tilde{k}_{1}^{1}-\tilde{k}%
_{2}^{2}\right) ^{2}\leq 0.
\end{equation}

This effective potential is negative for all surfaces and zero for flat
planes and spherical shells. Therefore, we conclude that being the surface
curved, in general, causes the particle to be bounded to the surface. The
strength of the binding potential depends on the square of the difference
between the principal curvatures. In other words, the larger $\left\vert 
\tilde{k}_{1}^{1}-\tilde{k}_{2}^{2}\right\vert $ the deeper the binding
potential.

\section{Monge parametrization}

Monge parametrization for a curved surface is given by%
\begin{equation}
\mathbf{r}=\left( x,y,H\left( x,y\right) \right)
\end{equation}%
in which the surface defined by 
\begin{equation}
z=H\left( x,y\right)
\end{equation}%
is described by $\left\{ x,y\right\} $, and $H\left( x,y\right) $ is called
the height function. The first fundamental form is given by%
\begin{equation}
h_{ab}=\left( 
\begin{array}{cc}
\mathbf{r}_{x}\mathbf{.r}_{x} & \mathbf{r}_{x}\mathbf{.r}_{y} \\ 
\mathbf{r}_{y}\mathbf{.r}_{x} & \mathbf{r}_{y}\mathbf{.r}_{y}%
\end{array}%
\right) =\left( 
\begin{array}{cc}
1+H_{x}^{2} & H_{x}H_{y} \\ 
H_{x}H_{y} & 1+H_{y}^{2}%
\end{array}%
\right)
\end{equation}%
with the corresponding line element on the surface%
\begin{equation}
ds^{2}=\left( 1+H_{x}^{2}\right) dx^{2}+\left( 1+H_{y}^{2}\right)
dy^{2}+2H_{x}H_{y}dxdy.
\end{equation}

The unit normal vector is obtained to be 
\begin{equation}
\mathbf{n}=\frac{\mathbf{r}_{x}\mathbf{\times r}_{y}}{\left\vert \mathbf{r}%
_{x}\mathbf{\times r}_{y}\right\vert }=\frac{\left( -H_{x},-H_{y},1\right) }{%
\sqrt{1+H_{x}^{2}+H_{y}^{2}}}
\end{equation}%
upon which, the second fundamental form is calculated as%
\begin{equation}
k_{ab}=-\mathbf{n}.\partial _{ab}\mathbf{r}=-\frac{1}{\sqrt{%
1+H_{x}^{2}+H_{y}^{2}}}\left( 
\begin{array}{cc}
H_{xx} & H_{xy} \\ 
H_{yx} & H_{yy}%
\end{array}%
\right) .
\end{equation}

Considering the inverse metric tensor 
\begin{equation}
h^{ab}=\frac{1}{1+H_{x}^{2}+H_{y}^{2}}\left( 
\begin{array}{cc}
1+H_{y}^{2} & -H_{x}H_{y} \\ 
-H_{x}H_{y} & 1+H_{x}^{2}%
\end{array}%
\right)
\end{equation}%
one finds%
\begin{equation}
k_{x}^{x}=-\frac{H_{xx}\left( 1+H_{y}^{2}\right) -H_{x}H_{y}H_{xy}}{\left(
1+H_{x}^{2}+H_{y}^{2}\right) ^{3/2}},
\end{equation}%
\begin{equation}
k_{y}^{y}=-\frac{H_{yy}\left( 1+H_{x}^{2}\right) -H_{x}H_{y}H_{xy}}{\left(
1+H_{x}^{2}+H_{y}^{2}\right) ^{3/2}},
\end{equation}%
\begin{equation}
k_{x}^{y}=\frac{H_{xx}H_{x}H_{y}-H_{xy}\left( 1+H_{x}^{2}\right) }{\left(
1+H_{x}^{2}+H_{y}^{2}\right) ^{3/2}}
\end{equation}%
and%
\begin{equation}
k_{y}^{x}=\frac{H_{yy}H_{x}H_{y}-H_{xy}\left( 1+H_{y}^{2}\right) }{\left(
1+H_{x}^{2}+H_{y}^{2}\right) ^{3/2}}.
\end{equation}

These imply%
\begin{equation}
K=k_{a}^{a}=-\frac{H_{xx}\left( 1+H_{y}^{2}\right) +H_{yy}\left(
1+H_{x}^{2}\right) -2H_{x}H_{y}H_{xy}}{\left( 1+H_{x}^{2}+H_{y}^{2}\right)
^{3/2}}
\end{equation}%
and%
\begin{equation}
K_{G}=\det k_{a}^{b}=\frac{H_{xx}H_{yy}-H_{xy}^{2}}{\left(
1+H_{x}^{2}+H_{y}^{2}\right) ^{2}}.
\end{equation}

Introducing, $\mathbf{\nabla }H=$ $\left( H_{x},H_{y}\right) $ one writes%
\begin{equation}
K=-\mathbf{\nabla .}\left( \frac{\mathbf{\nabla }H}{\sqrt{1+\left( \mathbf{%
\nabla }H\right) ^{2}}}\right)
\end{equation}%
and%
\begin{equation}
K_{G}=\frac{\det \partial ^{2}H}{\left( 1+\left( \mathbf{\nabla }H\right)
^{2}\right) ^{2}}
\end{equation}%
in which $\partial ^{2}H$ is called the Hessian of the function $H\left(
x,y\right) $ given by%
\begin{equation}
\partial ^{2}H=\left( 
\begin{array}{cc}
H_{xx} & H_{xy} \\ 
H_{yx} & H_{yy}%
\end{array}%
\right) .
\end{equation}

As a particular case, we may consider $H\left( x,y\right) =H\left( x\right) $
or $H\left( x,y\right) =H\left( y\right) $ upon which both result in the
same form of%
\begin{equation}
K=-\frac{H_{ii}}{\left( 1+H_{i}^{2}\right) ^{3/2}}
\end{equation}%
where $i=x,y$ and 
\begin{equation}
K_{G}=0.
\end{equation}

As we have mentioned at the beginning of this section, we are going to
consider the curved surface to be a small deviation from the flat surface
i.e.,%
\begin{equation}
\left\vert \mathbf{\nabla }H\right\vert \ll 1
\end{equation}%
such that%
\begin{equation}
K\simeq -\mathbf{\nabla }^{2}H=-Tr\partial ^{2}H
\end{equation}%
and%
\begin{equation}
K_{G}\simeq \det \partial ^{2}H.
\end{equation}

We note also that the determinant of the metric tensor i.e., 
\begin{equation}
g=1+\left( \mathbf{\nabla }H\right) ^{2}
\end{equation}%
is approximately 1 in this gauge i.e., 
\begin{equation}
g\simeq 1.
\end{equation}

Therefore, the resulting time-independent tangential Schr\"{o}dinger
equation of a particle confined on the surface $S$ is obtained to be%
\begin{equation}
-\frac{\hslash ^{2}}{2m}\nabla ^{2}\psi _{t}\left( x,y\right) +V_{S}\psi
_{t}\left( x,y\right) =E_{t}\psi _{t}\left( x,y\right)  \label{63}
\end{equation}%
where for the general configuration the effective potential is expressed as%
\begin{equation}
V_{S}=-\frac{\hslash ^{2}}{8m}\left( \left( \mathbf{\nabla .}\left( \frac{%
\mathbf{\nabla }H}{\sqrt{1+\left( \mathbf{\nabla }H\right) ^{2}}}\right)
\right) ^{2}-\frac{4\det \partial ^{2}H}{\left( 1+\left( \mathbf{\nabla }%
H\right) ^{2}\right) ^{2}}\right) .
\end{equation}

For the Monge gauge where $\left\vert \mathbf{\nabla }H\right\vert \ll 1$
one finds%
\begin{equation}
V_{S}\simeq -\frac{\hslash ^{2}}{2m}\left( \left( \frac{Tr\partial ^{2}H}{2}%
\right) ^{2}-\det \partial ^{2}H\right) 
\end{equation}%
which after simplification becomes%
\begin{equation}
V_{S}\simeq -\frac{\hslash ^{2}}{8m}\left( \left( H_{xx}-H_{yy}\right)
^{2}+4H_{xy}^{2}\right) .
\end{equation}%
Moreover, for the case where $H$ is only a function of one coordinate, the
latter becomes ($i=x,y$)%
\begin{equation}
V_{S}=-\frac{\hslash ^{2}}{8m}\frac{H_{ii}^{2}}{\left( 1+H_{i}^{2}\right)
^{3}}
\end{equation}%
which in the gauge where $H_{i}^{2}\ll 1$ it reduces to%
\begin{equation}
V_{S}\simeq -\frac{\hslash ^{2}}{8m}H_{ii}^{2}.
\end{equation}

\subsection{Monge gauge in polar coordinate with radial symmetry}

Considering an axial symmetric surface defined by 
\begin{equation}
z\left( \rho ,\theta \right) =H\left( \rho \right)
\end{equation}%
in cylindrical coordinate system $\left\{ \rho ,\theta ,z\right\} $,
admitting radial symmetry, one obtains%
\begin{equation}
g_{ab}=\left( 
\begin{array}{cc}
1+H_{\rho }^{2} & 0 \\ 
0 & \rho ^{2}%
\end{array}%
\right)
\end{equation}%
with the line element%
\begin{equation}
ds^{2}=\left( 1+H_{\rho }^{2}\right) d\rho ^{2}+\rho ^{2}d\theta ^{2}.
\end{equation}

The unit normal vector and the second fundamental form are obtained to be%
\begin{equation}
n_{a}=\frac{1}{\sqrt{1+H_{\rho }^{2}}}\left( -H_{\rho },0,1\right) 
\end{equation}%
and 
\begin{equation}
k_{ab}=\left( 
\begin{array}{cc}
\frac{-H_{\rho \rho }}{\sqrt{1+H_{\rho }^{2}}} & 0 \\ 
0 & \frac{-H_{\rho }}{\sqrt{1+H_{\rho }^{2}}}\rho 
\end{array}%
\right) .
\end{equation}%
Consequently, we calculate 
\begin{equation}
k_{a}^{b}=\left( 
\begin{array}{cc}
\frac{-H_{\rho \rho }}{\left( 1+H_{\rho }^{2}\right) ^{3/2}} & 0 \\ 
0 & \frac{-H_{\rho }}{\sqrt{1+H_{\rho }^{2}}}\frac{1}{\rho }%
\end{array}%
\right) .
\end{equation}%
Having $k_{a}^{b}$ diagonal, the principal curvatures are given by%
\begin{equation}
\tilde{k}_{\rho }^{\rho }=\frac{-H_{\rho \rho }}{\left( 1+H_{\rho
}^{2}\right) ^{3/2}}  \label{75}
\end{equation}%
and%
\begin{equation}
\tilde{k}_{\theta }^{\theta }=\frac{-H_{\rho }}{\sqrt{1+H_{\rho }^{2}}}\frac{%
1}{\rho }.  \label{76}
\end{equation}%
Using (\ref{75}) and (\ref{76}), the time independent tangent Schr\"{o}%
dinger equation on the surface becomes%
\begin{equation}
-\frac{\hslash ^{2}}{2m}\frac{1}{\sqrt{h}}\partial _{a}\left( \sqrt{h}%
g^{ab}\partial _{b}\right) \psi _{t}\left( \rho ,\theta \right) -\frac{%
\hslash ^{2}}{8m}\left( \frac{H_{\rho \rho }}{\left( 1+H_{\rho }^{2}\right)
^{3/2}}-\frac{H_{\rho }}{\sqrt{1+H_{\rho }^{2}}}\frac{1}{\rho }\right)
^{2}\psi _{t}\left( \rho ,\theta \right) =E_{t}\psi _{t}\left( \rho ,\theta
\right) 
\end{equation}%
in which $h=\rho ^{2}\left( 1+H_{\rho }^{2}\right) .$ In the small curvature
limit where $\left\vert H_{\rho }\right\vert \ll 1$, we find 
\begin{equation}
V_{S}=-\frac{\hslash ^{2}}{8m}\left( H_{\rho \rho }-\frac{H_{\rho }}{\rho }%
\right) ^{2}=-\frac{\hslash ^{2}}{8m}\rho ^{2}\left[ \partial _{\rho }\left( 
\frac{H_{\rho }}{\rho }\right) \right] ^{2}.
\end{equation}

\section{Catenary Surface}

Consider an infinite flat plane that is bent into a Catenary surface defined
by%
\begin{equation}
H\left( x,y\right) =H\left( x\right) =a\cosh \left( \frac{x}{a}\right) 
\end{equation}%
where $a$ is a positive real constant. The corresponding Schr\"{o}dinger
equation of a particle confined on this surface is given by (\ref{63}).
Corresponding to the curved surface, the geometric potential, the metric
tensor, and the determinant of the metric tensor are obtained to be%
\begin{equation}
V_{S}=-\frac{\hslash ^{2}}{8m}\frac{H_{xx}^{2}}{\left( 1+H_{x}^{2}\right)
^{3}}=-\frac{\hslash ^{2}}{8m}\frac{1}{a^{2}\cosh ^{4}\left( \frac{x}{a}%
\right) },
\end{equation}%
\begin{equation}
g_{ab}=diag\left[ 1+\sinh ^{2}\left( \frac{x}{a}\right) ,1\right] 
\end{equation}%
and 
\begin{equation*}
g=1+H^{\prime }(x)^{2}=\cosh ^{2}\left( \frac{x}{a}\right) ,
\end{equation*}%
respectively. In Fig. 2 we plot the Catenary surface together with the
corresponding effective potential $V_{S}/\left( \frac{\hslash ^{2}}{m}%
\right) .$

\begin{figure}[tbp]
\caption{\protect\includegraphics[scale=0.6]{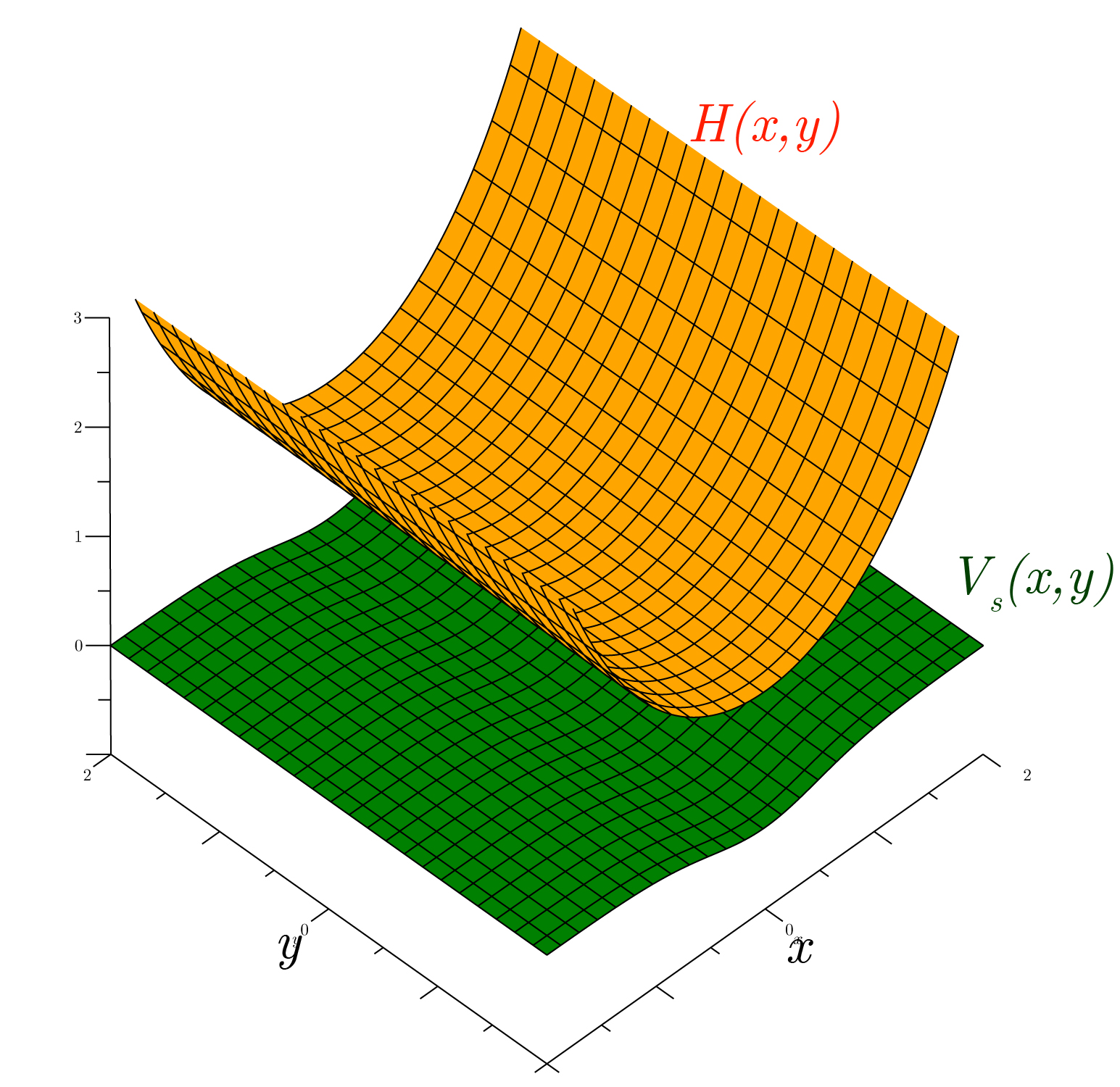}}
The Catenary surface and the corresponding geometric potential on the
surface, observed by the particle.
\end{figure}
Considering $\psi _{t}\left( x,y\right) =X\left( x\right) Y\left( y\right) $
one obtains%
\begin{equation}
-\frac{\hslash ^{2}}{2m}\frac{1}{\cosh \left( \frac{x}{a}\right) }\frac{d}{dx%
}\left( \frac{1}{\cosh \left( \frac{x}{a}\right) }\frac{d}{dx}\right)
X\left( x\right) -\frac{\hslash ^{2}}{8m}\frac{1}{a^{2}\cosh ^{4}\left( 
\frac{x}{a}\right) }X\left( x\right) =E_{x}X\left( x\right)
\end{equation}%
and%
\begin{equation}
-\frac{\hslash ^{2}}{2m}Y^{\prime \prime }\left( y\right) =E_{y}Y\left(
y\right) .
\end{equation}

Let's introduce $\alpha =\frac{2mE_{x}}{\hslash ^{2}}$ and $\beta =\frac{%
2mE_{y}}{\hslash ^{2}}$ upon which (80) and (81) become 
\begin{equation}
-\frac{1}{\cosh \left( \frac{x}{a}\right) }\frac{d}{dx}\left( \frac{1}{\cosh
\left( \frac{x}{a}\right) }\frac{d}{dx}\right) X\left( x\right) -\frac{1}{4}%
\frac{1}{a^{2}\cosh ^{4}\left( \frac{x}{a}\right) }X\left( x\right) =\alpha
X\left( x\right)   \label{84}
\end{equation}%
and%
\begin{equation}
Y^{\prime \prime }\left( y\right) +\beta Y\left( y\right) =0,
\end{equation}%
respectively. To solve the $x$-component of the field equations, we
introduce a change of variable in the following form%
\begin{equation}
q=\sinh \left( \frac{x}{a}\right) 
\end{equation}%
which after some manipulation (\ref{84}) becomes%
\begin{equation}
-X^{\prime \prime }\left( q\right) -\frac{1}{4\left( 1+q^{2}\right) ^{2}}%
X\left( q\right) =\mathcal{E}X\left( q\right) ,  \label{87}
\end{equation}%
in which $\mathcal{E}=a^{2}\alpha $. The latter equation describes a
one-dimensional quantum particle in $q$-space which undergoes a binding
potential of the form%
\begin{equation}
V\left( q\right) =-\frac{1}{4\left( 1+q^{2}\right) ^{2}}.
\end{equation}%
Further, to solve (\ref{87}) we introduce another change of variable given
as $z=-q^{2}$ and $X\left( q\right) =\left( 1-z\right) ^{\lambda }w\left(
z\right) $ with $\lambda =\frac{1}{2}+\frac{\sqrt{5}}{4},$ upon which (\ref%
{87}) becomes%
\begin{equation}
w^{\prime \prime }+\left( \alpha +\frac{\beta +1}{z}+\frac{\gamma +1}{z-1}%
\right) w^{\prime }+\left( \frac{\mu }{z}+\frac{\nu }{z-1}\right) w=0
\label{89}
\end{equation}%
where%
\begin{equation}
\alpha =0,\beta =-\frac{1}{2},\gamma =\frac{\sqrt{5}}{2},\mu =-\frac{4E+5+2%
\sqrt{5}}{16}\text{ and }\nu =\frac{1}{8}\left( \frac{5}{2}+\sqrt{5}\right) .
\end{equation}%
Eq. (\ref{89}) is the so-called Confluent Heun Differential Equation (CHDE)
whose solution is given by%
\begin{equation}
w\left( z\right) =C_{1}HeunC\left( \alpha ,\beta ,\gamma ,\delta ,\eta
,z\right) +C_{2}z^{-\beta }HeunC\left( \alpha ,-\beta ,\gamma ,\delta ,\eta
,z\right) .
\end{equation}%
Herein, $C_{1}$ and $C_{2}$ are two integration constants and $\delta =\mu
+\nu -\frac{\alpha \left( 2+\gamma +\beta \right) }{2}=-\frac{\mathcal{E}}{4}
$ and $\eta =\frac{\left( \beta +1\right) \left( \alpha -\gamma \right)
-\beta }{2}-\mu =\frac{\mathcal{E}}{4}+\frac{9}{16}.$ Hence, the general
solution of the Schr\"{o}dinger equation becomes%
\begin{multline}
X\left( q\right) =C_{1}\left( 1+q^{2}\right) ^{\frac{1}{2}+\frac{\sqrt{5}}{4}%
}HeunC\left( 0,-\frac{1}{2},\frac{\sqrt{5}}{2},-\frac{\mathcal{E}}{4},\frac{%
\mathcal{E}}{4}+\frac{9}{16},-q^{2}\right) + \\
C_{2}\left( 1+q^{2}\right) ^{\frac{1}{2}+\frac{\sqrt{5}}{4}}\sqrt{q^{2}}%
HeunC\left( 0,\frac{1}{2},\frac{\sqrt{5}}{2},-\frac{\mathcal{E}}{4},\frac{%
\mathcal{E}}{4}+\frac{9}{16},-q^{2}\right) .
\end{multline}%
Our detailed numerical analysis reveals that there exists only one bound
state with $\mathcal{E}=-0.02892460$, $C_{2}=0,$ and $C_{1}=\frac{1}{\sqrt{%
6.7406}}.$ Hence, the normalized wave function is found to be 
\begin{equation}
X\left( q\right) =\frac{1}{\sqrt{6.7406}}\left( 1+q^{2}\right) ^{\frac{1}{2}+%
\frac{\sqrt{5}}{4}}HeunC\left( 0,-\frac{1}{2},\frac{\sqrt{5}}{2},-\frac{%
\mathcal{E}}{4},\frac{\mathcal{E}}{4}+\frac{9}{16},-q^{2}\right) 
\end{equation}%
which together with the potential $V\left( q\right) =-\frac{1}{4\left(
1+q^{2}\right) ^{2}}$ are plotted in Fig. 3. This figure reveals that the
particle is localized around the deep of the surface where the total
curvature is maximum.

\begin{figure}[tbp]
\caption{\protect\includegraphics[scale=0.6]{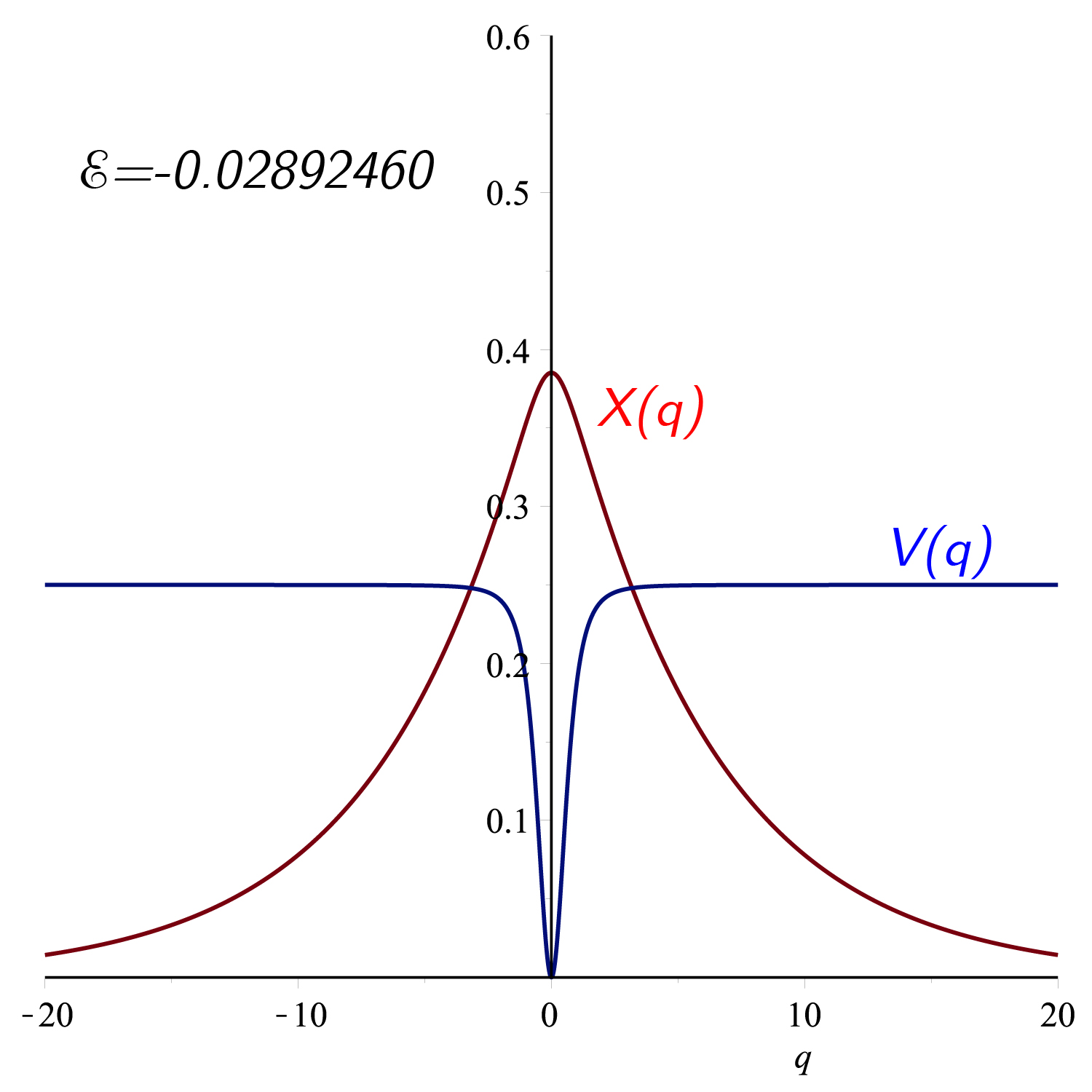}}
The normalized wave function of the particle confined to move on a Catenary
surface together with the pure geometrical potential the particle enconteres
in $q$-space.
\end{figure}

\section{Paraboloid of Revolution}

Concerning the results of the polar coordinate with radial symmetry, we
consider the surface of the paraboloid of revolution which is defined as%
\begin{equation}
H\left( \rho \right) =a\left( \frac{\rho }{a}\right) ^{2}
\end{equation}%
where $a$ is a constant parameter with the dimension of length. In Fig. 4 we
plot the paraboloid of revolution for $a=1.$ The line element on the surface
of the paraboloid of revolution is obtained to be%
\begin{equation}
ds^{2}=\left( 1+\frac{\rho ^{2}}{a^{2}}\right) d\rho ^{2}+\rho ^{2}d\theta
^{2}.
\end{equation}%
\begin{figure}[tbp]
\caption{\protect\includegraphics[scale=0.6]{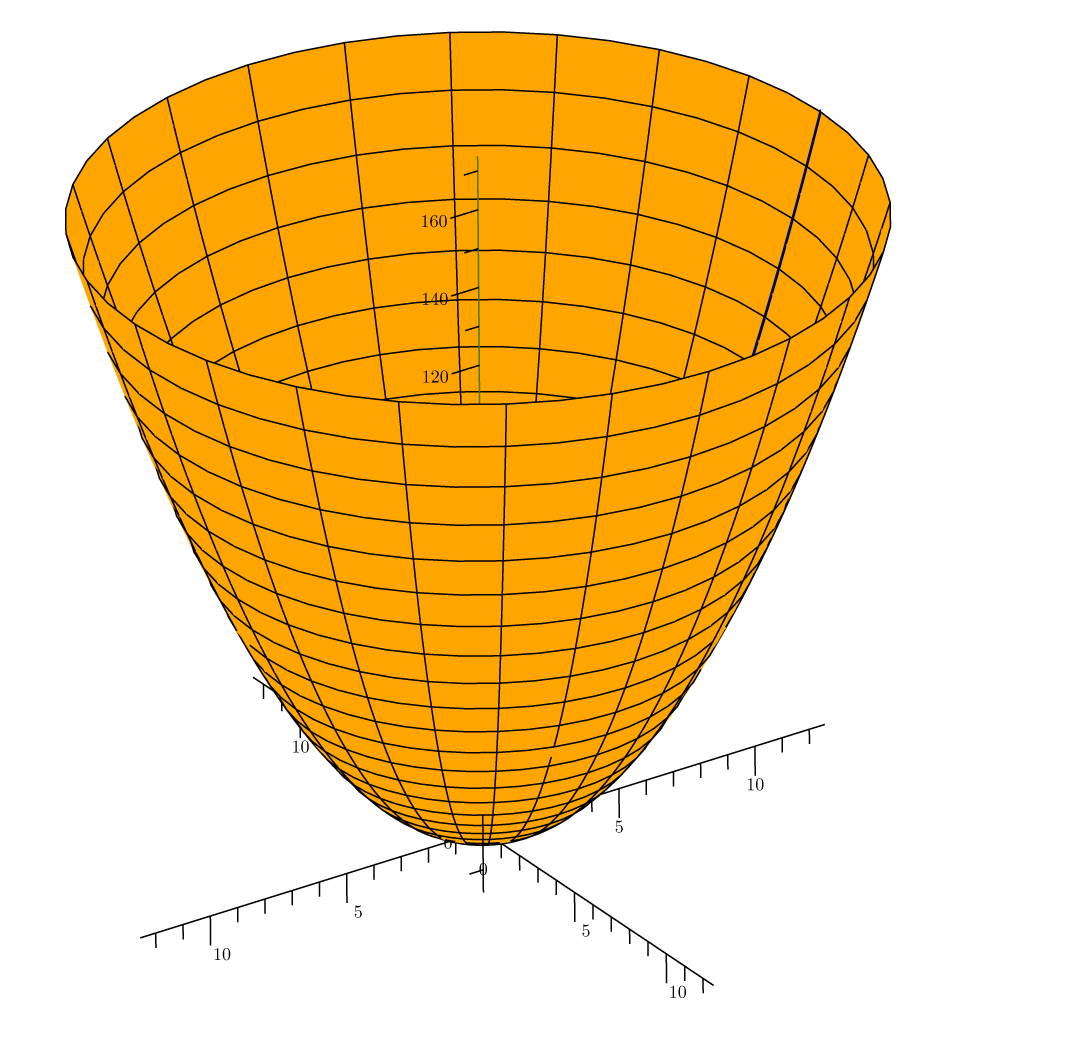}}
The graph of the Paraboloid of Revolution with $a=1$.
\end{figure}
\begin{figure}[tbp]
\caption{\protect\includegraphics[scale=0.6]{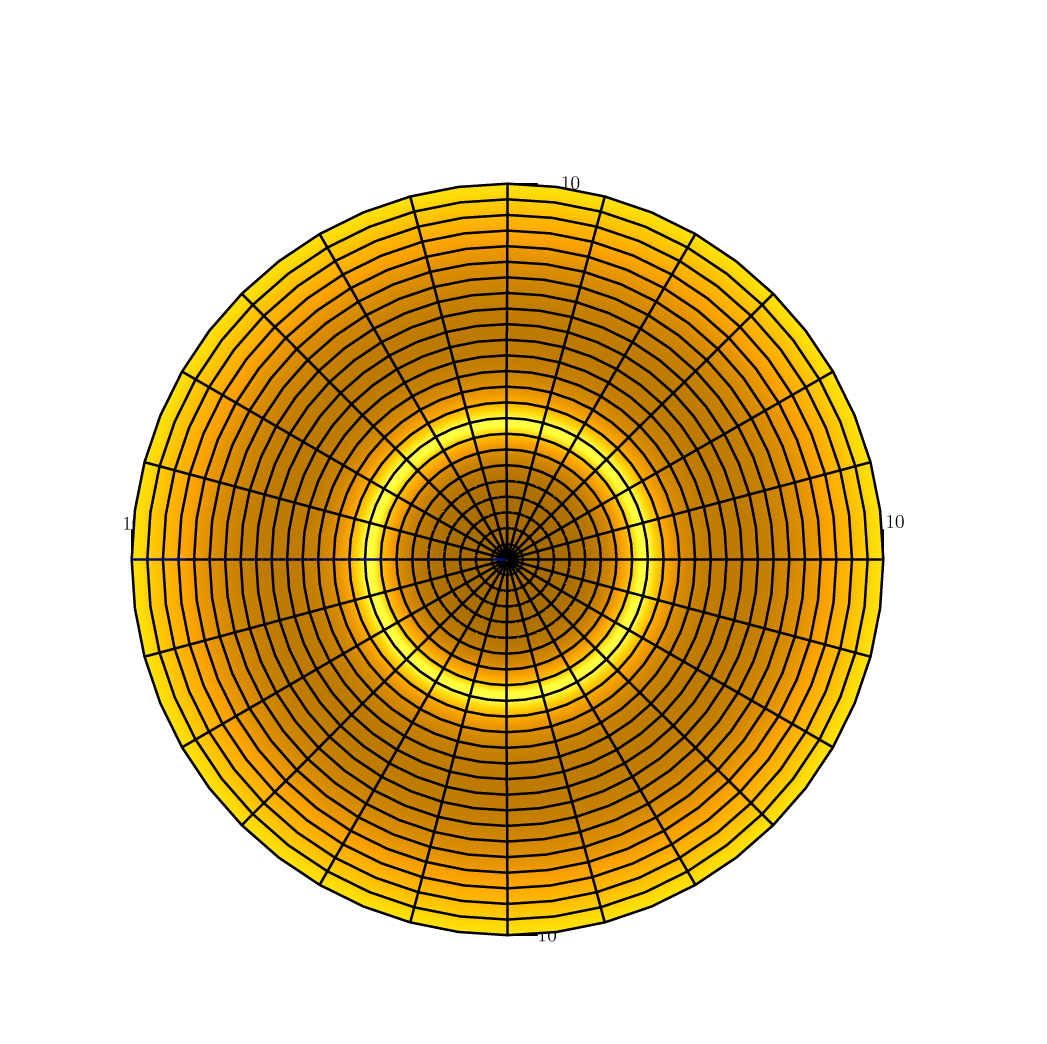}}
The wave function $\rho \psi _{t}\left( \rho ,\theta \right) $ with respect
to $\rho $ and $\theta .$ The brightest circle implies the maximum
probability of finding the particle on the surface.
\end{figure}

\begin{figure}[tbp]
\caption{\protect\includegraphics[scale=0.6]{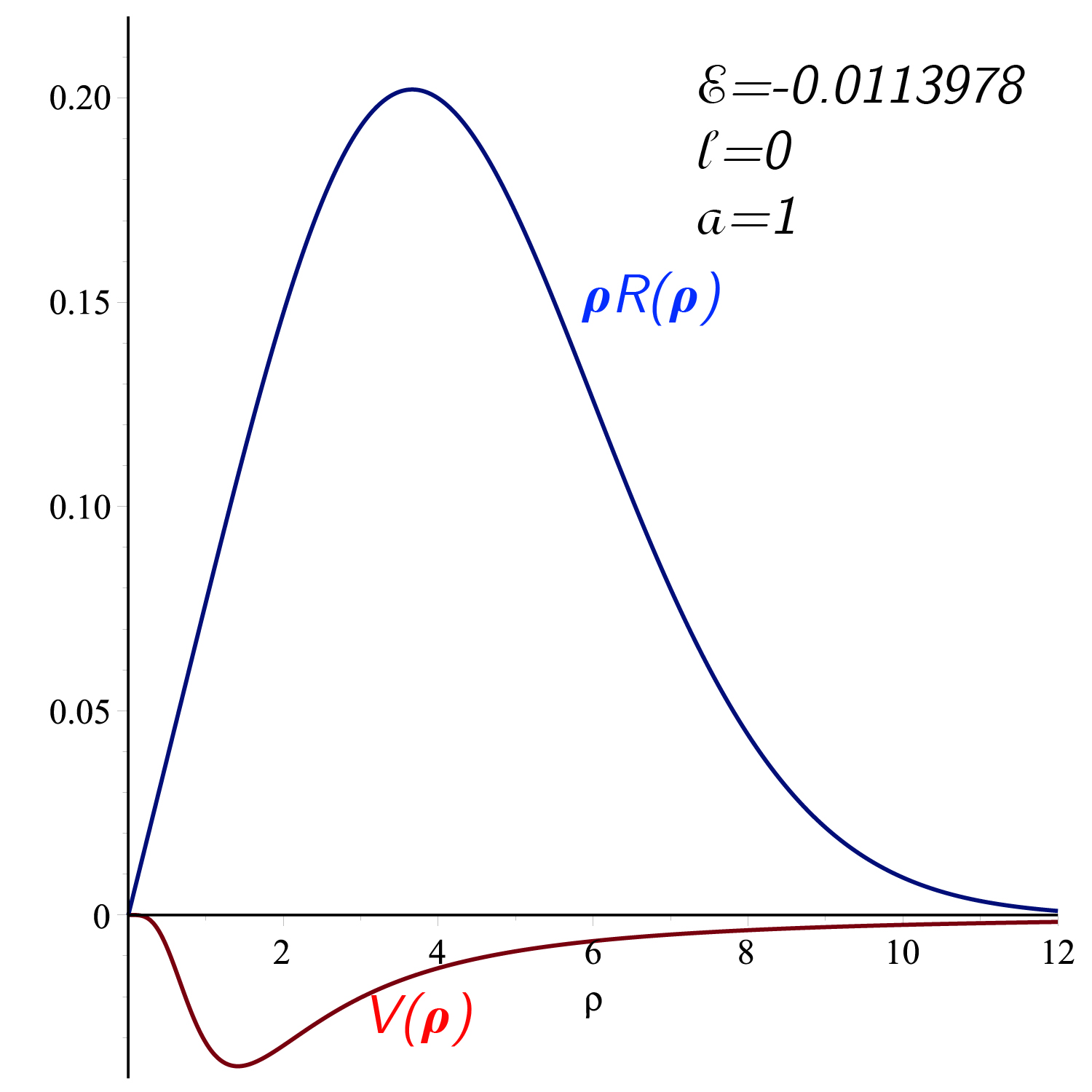}}
The wave function $\rho R\left( \rho \right) $ versus the potential $%
V_S(\rho)$. The maximum probability doesn't correspond to the minimum of the
potential.
\end{figure}

\begin{figure}[tbp]
\caption{\protect\includegraphics[scale=0.6]{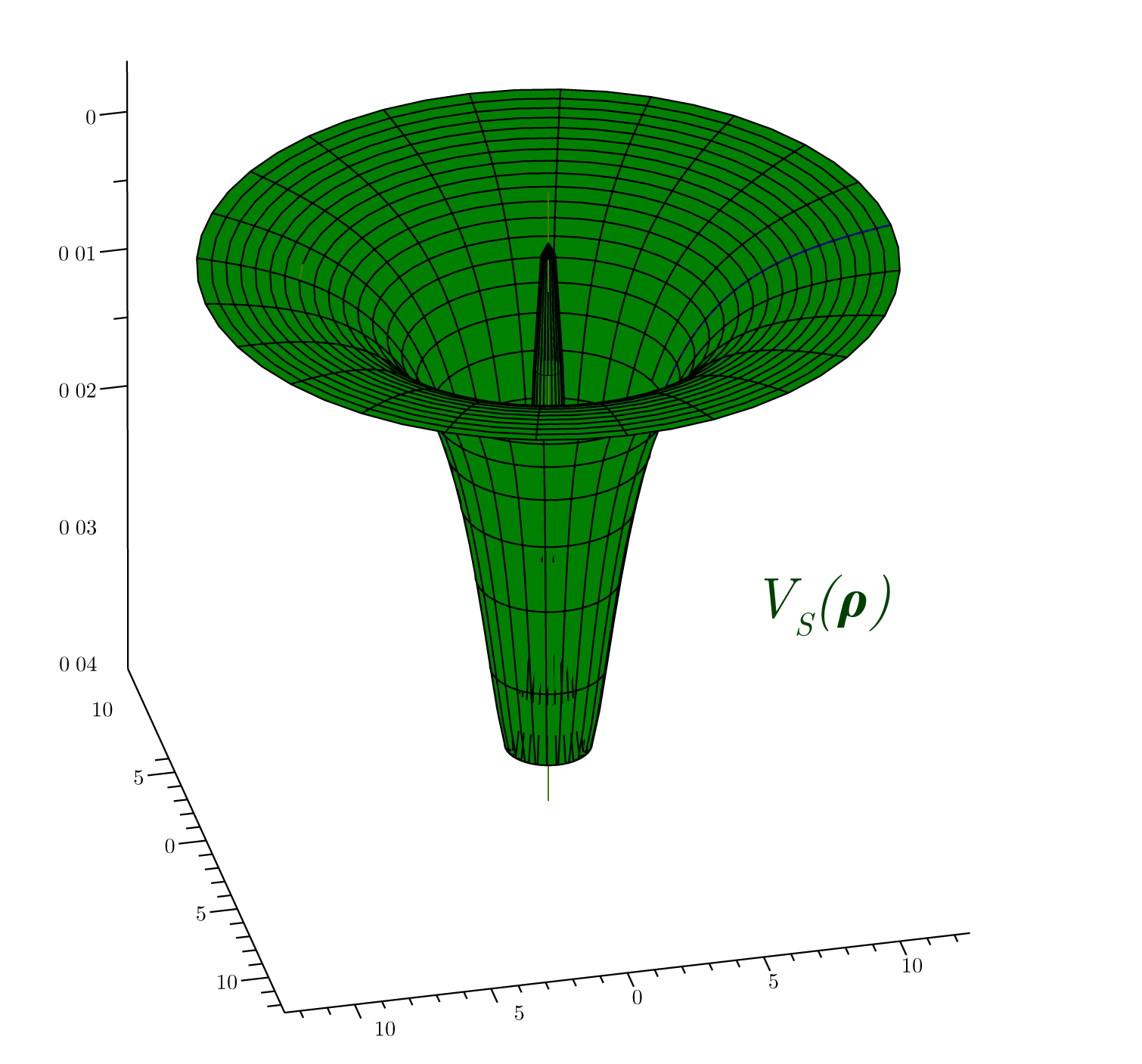}}
The three dimensional shape of the pure geometric potential experienced by
the quantum particle moving on the paraboloid of revolution.
\end{figure}
The two dimensional Schr\"{o}dinger equation on the surface is given by 
\begin{equation}
-\frac{\hslash ^{2}}{2m}\frac{1}{\sqrt{h}}\partial _{a}\left( \sqrt{h}%
g^{ab}\partial _{b}\right) \psi _{t}\left( \rho ,\theta \right) +V_{S}\left(
\rho \right) \psi _{t}\left( \rho ,\theta \right) =E_{t}\psi _{t}\left( \rho
,\theta \right) 
\end{equation}%
in which 
\begin{equation}
V_{S}\left( r\right) =-\frac{\hslash ^{2}}{8m}\frac{\rho ^{4}}{\left(
a^{2}+\rho ^{2}\right) ^{3}},
\end{equation}%
\begin{equation}
h=\det g_{ab}=\rho ^{2}\left( 1+\frac{\rho ^{2}}{a^{2}}\right) 
\end{equation}%
and 
\begin{equation}
g^{ab}=diag\left[ \frac{1}{1+\frac{\rho ^{2}}{a^{2}}},\frac{1}{\rho ^{2}}%
\right] .
\end{equation}%
Applying the separating method one obtains%
\begin{equation}
\psi _{t}\left( \rho ,\theta \right) =R\left( \rho \right) e^{i\ell \theta }
\end{equation}%
with $\ell =0,\pm 1,\pm 2,...$ and $R\left( r\right) $ satisfying%
\begin{equation}
-R^{\prime \prime }-\frac{a^{2}}{\rho \left( a^{2}+\rho ^{2}\right) }%
R^{\prime }+\left( \frac{\ell ^{2}\left( a^{2}+\rho ^{2}\right) }{a^{2}\rho
^{2}}-\frac{\rho ^{4}}{4a^{2}\left( a^{2}+\rho ^{2}\right) ^{2}}\right) R=%
\frac{a^{2}+\rho ^{2}}{a^{2}}\mathcal{E}R
\end{equation}%
where $\mathcal{E}=\frac{2mE_{t}}{\hslash ^{2}}.$ Without going into the
details, the general solution of the Schr\"{o}dinger equation is given in
terms of the Confluent Heun function, expressed as 
\begin{multline}
R\left( \rho \right) =C_{1}\rho ^{\ell }\left( a^{2}+\rho ^{2}\right) ^{%
\frac{3+\sqrt{10}}{4}}e^{-\frac{\sqrt{\left\vert \mathcal{E}\right\vert }}{2a%
}\rho ^{2}}HeunC\left( a\sqrt{\left\vert \mathcal{E}\right\vert },\ell ,%
\frac{\sqrt{10}}{2},-\frac{1}{16}+\frac{\ell ^{2}}{4}-\frac{\mathcal{E}a^{2}%
}{4},\frac{3-\ell ^{2}+\mathcal{E}a^{2}}{4},-\frac{\rho ^{2}}{a^{2}}\right) +
\label{102} \\
C_{2}\rho ^{-\ell }\left( a^{2}+\rho ^{2}\right) ^{\frac{3+\sqrt{10}}{4}}e^{-%
\frac{\sqrt{\left\vert \mathcal{E}\right\vert }}{2a}\rho ^{2}}HeunC\left( a%
\sqrt{\left\vert \mathcal{E}\right\vert },-\ell ,\frac{\sqrt{10}}{2},-\frac{1%
}{16}+\frac{\ell ^{2}}{4}-\frac{\mathcal{E}a^{2}}{4},\frac{3-\ell ^{2}+%
\mathcal{E}a^{2}}{4},-\frac{\rho ^{2}}{a^{2}}\right) ,
\end{multline}%
in which $C_{1}$ and $C-2$ are the integration constants. From the solution (%
\ref{102}) we see that with $\ell \geq 0$ the first solution is regular at
the origin while for $\ell \leq 0$ the second solution coincides with the
first solution (with $\ell \geq 0)$. Knowing also that the sign of $\ell $
doesn't make any change in the solution, we set $\ell \geq 0$ and eliminate
the second solution as it is irregular at the origin. The square
integrability of the solution implies%
\begin{equation}
\int\nolimits_{0}^{2\pi }\int\nolimits_{0}^{\infty }\left\vert R\left( \rho
\right) e^{i\ell \theta }\right\vert ^{2}\rho ^{2}d\rho d\theta =1.
\end{equation}%
Our numerical calculations revealed that for the ground state where $\ell =0$
the binding energy is given by $\mathcal{E}=-0.0113978$ and $C_{1}=\frac{1}{%
\sqrt{2\pi }\sqrt{27.268}}.$ In Fig. 5 we plot $\rho \psi _{t}\left( \rho
,\theta \right) $ in terms of $\rho $ and $\theta .$ The bright circle is
the pick of the wave function indicating the maximum probability radius. In
Fig. 6 we plot the normalized wave function $\rho R\left( \rho \right) $ and
the potential $V_{S}\left( \rho \right) $ in terms of $\rho $ for $\ell =0$
and $a=1.$ In Fig. 7 we plot a three-dimensional picture of the potential
which the particle on the surface observes. We comment finally that for $%
\ell \geq 0$ there is no bound state solution satisfying the boundary
conditions.

\section{Conclusion}

The nonrelativistic quantum particle confined to a curved surface has been
revisited in a more familiar notation and more details. We have studied
explicitly the case of the Monge parametrization in two different coordinate
systems, i.e., Cartesian and Polar coordinates. We have shown that in the
Cartesian coordinate system the geometric effective potential for the small
perturbation is simply given by $V_{S}\simeq -\frac{\hslash ^{2}}{8m}\left(
\left( H_{xx}-H_{yy}\right) ^{2}+4H_{xy}^{2}\right) $ which depends on the
second derivatives of the height function $H\left( x,y\right) .$ The
effective Schr\"{o}dinger equation i.e., 
\begin{equation}
-\frac{\hslash ^{2}}{2m}\nabla ^{2}\psi _{t}\left( x,y\right) -\frac{\hslash
^{2}}{8m}\left( \left( H_{xx}-H_{yy}\right) ^{2}+4H_{xy}^{2}\right) \psi
_{t}\left( x,y\right) =E_{t}\psi _{t}\left( x,y\right)
\end{equation}%
with $\nabla ^{2}=\frac{\partial ^{2}}{\partial x^{2}}+\frac{\partial ^{2}}{%
\partial y^{2}}$ is significantly simpler than the original one. In many
cases where the deviation from a flat surface is small, we believe that this
equation is a very good and acceptable approximation. In the last part of
the paper, we studied two interesting curved surfaces, particularly a
Catenoid and a paraboloid of revolution. We solved the corresponding Schr%
\"{o}dinger equation of a particle confined on these surfaces without any
external potential. We found only one possible\ bound state for each surface
which localizes the particle around the deep of the point of maximum
curvature. The exact normalized wave functions with the potentials and the
surfaces have been displayed in a number of figures.

\end{document}